\documentclass{iau} 
\usepackage{graphicx}

\title[Dust precursors in O-rich stars] %% give here short title %%
{Molecular dust precursors in envelopes of oxygen-rich AGB stars and red supergiants}

\author[Tomek Kami\'nski]   %% give here short author list %%
{Tomasz Kami\'nski}

\affiliation{ 
	Harvard-Smithsonian Center for Astrophysics, 
	60 Garden Street,
	Cambridge, MA 02138\\ Submillimeter Array Fellow, email: {\tt tkaminsk@cfa.harvard.edu}}
\pubyear{2018}
\volume{343}  %% insert here IAU Symposium No.
\jname{Why Galaxies Care About AGB Stars: A Continuing Challenge through Cosmic Time}
\editors{F. Kerschbaum, H. Olofsson \& M. Groenewegen, eds.}
\begin{document}

\maketitle

\begin{abstract}
Condensation of circumstellar dust begins with formation of molecular clusters close to the stellar photosphere. These clusters are predicted to act as condensation cores at lower temperatures and allow efficient dust formation farther away from the star. Recent observations of metal oxides, such as AlO, AlOH, TiO, and TiO$_2$, whose emission can be traced at high angular resolutions with ALMA, have allowed first observational studies of the condensation process in oxygen-rich stars. We are now in the era when depletion of gas-phase species into dust can be observed directly. I review the most recent observations that allow us to identify gas species involved in the formation of inorganic dust of AGB stars and red supergiants. I also discuss challenges we face in interpreting the observations, especially those related to non-equilibrium gas excitation and the high complexity of stellar atmospheres in the dust-formation zone. 
\end{abstract}

\firstsection % if your document starts with a section,
\section{Dust formation and seed particles in O-rich stars}
Galaxies certainly care about AGB stars --- in the local Universe, galaxies owe huge amounts of dust to these unexhaustive factories of cosmic solids. Inorganic dust, that is, the silicate and alumina dust, is formed primarily in oxygen-rich environments of M-type AGB stars and red supergiants. Despite the crucial role of dust in a broad range of astrophysical phenomena, the formation mechanisms of stardust are poorly understood, mainly because they involve complex problems of dynamic stellar atmospheres and shock-driven chemistry. Observations readily indicate the presence of hot inorganic dust close to stellar photospheres which points to efficient formation of first solids, or \textit{seeds}, out of molecular gas at temperatures of $\sim$1100--1700\,K. I describe here observational efforts to identify these gas-phase species and physio-chemical processes involved in the nucleation process close to a pulsating atmosphere. The complexities of the dust nucleation processes in circumstellar envelopes are described in detail in \cite{bookGS}. 

\section{How can we identify the dust precursors?}\label{S2}
%% theory, name poterntial precursors
{\underline{\it Theoretical predictions.}}
The quest to identify inorganic seed particles was initiated and is continued by theoretical studies. The strongest requirement for seed particles is that they have to form from gas-phase molecules that are abundant and have high nucleation rates. For instance, \cite{Jeong2003} considered Fe, SiO, SiO$_2$, TiO and TiO$_2$, and Al$_2$O$_3$ as dust precursors but only TiO and TiO$_2$ appeared to have high enough nucleation rates in the relevant physical conditions (see also \cite{GailSedlmayr1998} and \cite{SharpHuebner1990}). Given the low abundance of titanium in material of cosmic composition, efficient seed formation from titanium oxides requires nearly all titanium to be depleted into these first solids, which is a questionable but observationally verifiable assumption. From the theoretical standpoint, the role of the hot silicon oxides in dust formation is debated and was recently reinstated after new laboratory measurements \cite{Gail2013}. Alumina oxides have long been strong candidates for the seed particles, including in recent studies of \cite{Cherchneff2013}, \cite{Gobrecht2016}, and \cite{hofner2016} (see also \cite{DellAgli}). A wide range of other species were considered (e.g. \cite{Ferrarotti2002} and \cite{Plane2013}), but most studies strongly favor Al$_2$O$_3$ and TiO$_2$ as the best candidates. Can we verify these theoretical predictions through  observations (or experiments)?

% presolar grains
{\underline{\it Presolar grains.}}
Laboratory measurements of pre-solar grains deliver important clues on the nucleation process. In particular, a handful of presolar grains were found to contain titanium oxides. Many more grains were identified to contain alumina solids.  There are cases where alumina cores are surrounded by silicate mantles (\cite{Nittler2008}), in accord with some of the theoretical expectations. However, the studies of presolar grains have their biases and  limitations, for instance, related to the location of the Solar System and analysis methods that give us access to the largest grains only. The field is advancing with an increasing number of analyzed grains of different types (\cite{Takigawa2018,Leitner}).  

% infrared observations
{\underline{\it Mid-infared spectroscopy.}}
One way to study the nucleation process is to directly observe the first hot solids in the closest M-type stars. To distinguish the hot newly-formed dust from the bulk of warm dust produced at larger radii from the star, such studies require very high angular resolutions available only to interferometers. Different chemical types and crystalline/amorphous forms of dust can be investigated through broad spectral features, located mainly in the mid-infrared (MIR). One such study was conducted by \cite{Karovicova2013} using the MIDI instrument on Very Large Telescope Interferometer (VLTI). For the M-type mira GX Mon, they found that alumina dust particles must be present at radii as close as 2.1\,$R_{\star}$ from the star and silicates are only present at distances greater than 5.5\,$R_{\star}$ (where $R_{\star}$ denotes the radius of the star). This stratification is consistent with what we would expect if alumina dust provides the condensation cores for silicate dust formation. Unfortunately, such studies are still very rare (\cite{Norris2012,Khouri2015}) and MIR spectroscopy does not always lead to unique identification of the carriers (\cite{Decin2017}). 

\begin{figure}[t!]
% \vspace*{-2.0 cm}
\begin{center}
 \fbox{
 \includegraphics[width=4in]{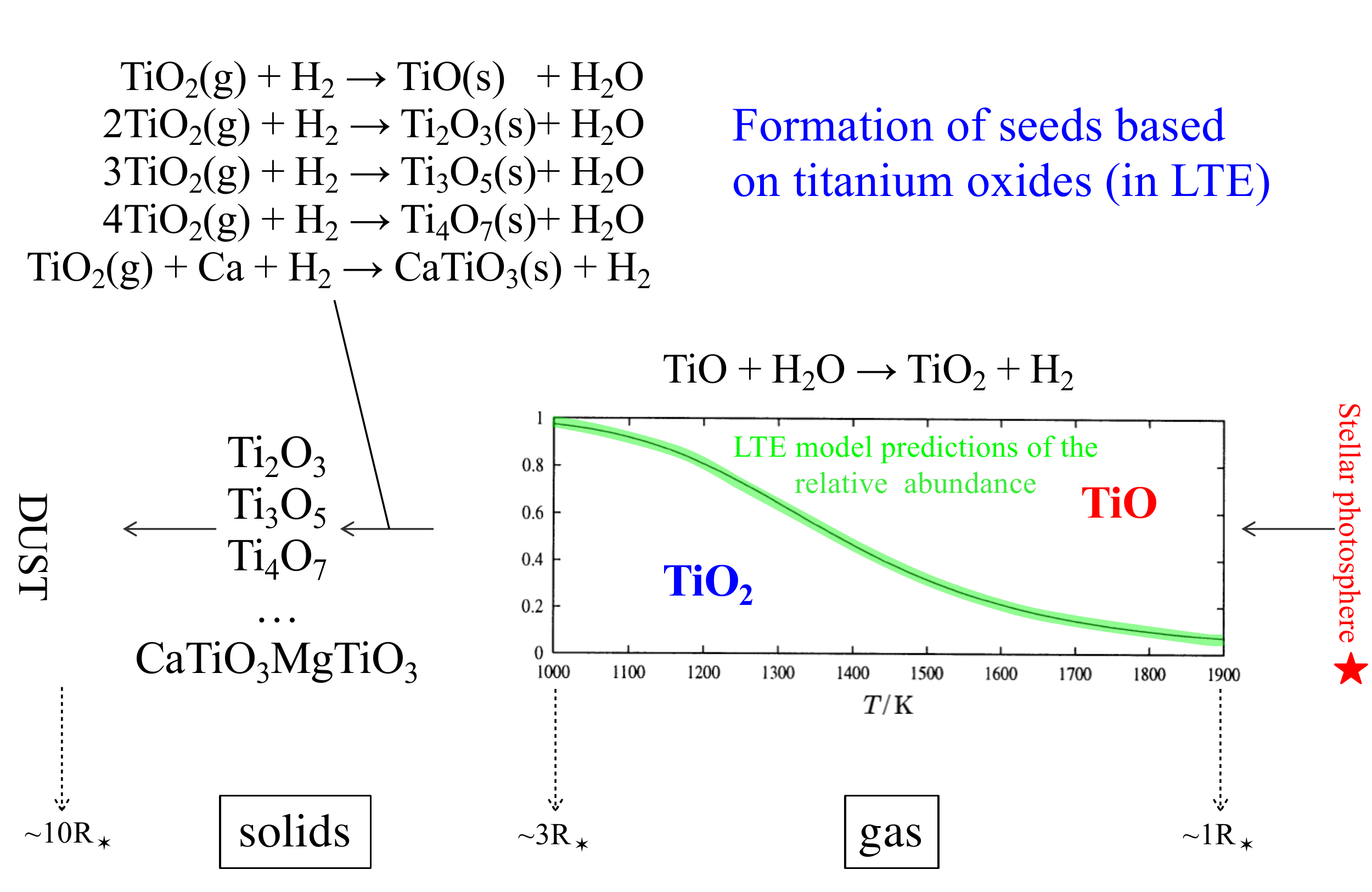} }
% \vspace*{-1.0 cm}
 \caption{Illustration of seed and dust formation starting from titanium oxides. The sketch shows the stellar photosphere on the right side and the schematic representation of the different chemical processes taking place at different distances from the star. The distances increase to the left and are not in scale. See text for details. Figure is based on \cite{GailSedlmayr1998}.}
   \label{fig1}
\end{center}
\end{figure}

% depletion
{\underline{\it Gas depletion.}}
In this contribution, we explore yet another observational possibility to study nucleation, that is, through signatures of depletion of the gas-phase precursors of the first solids that condense. The idea is simple: we should look for simple molecules whose abundance drops down in the dust-formation zone owing to their nucleation and incorporation into solids. This, in principle, should allow us to figure out what kind of compounds are involved in the process and how effective the nucleation is. Let us consider an example that we schematically illustrate in Fig.\,\ref{fig1}. Assuming the fist solids are formed from titanium oxides, dust formation must start with the simplest molecule, TiO. As a plethora of studies that go back to the earliest days of optical stellar spectroscopy show, titanium monoxide exists already in the cool photospheres of M-type stars (say, at 2000--3000\,K). Once it is transported to the circumstellar material (whatever the transport process may be) it reacts with water, very abundant in these O-rich environments, forming TiO$_2$. The parcel of gas is moved away from the photosphere and, as it becomes cooler, more and more of TiO is converted into gaseous TiO$_2$. In the LTE calculations of \cite{GailSedlmayr1998}, all TiO is oxidized to TiO$_2$ at a temperature of about 1000\,K or distances equivalent to 3\,$R_{\star}$. At even lower temperatures, gaseous TiO$_2$ (and TiO) may react with H$_2$ to form first simple solids which may stick to each other and form clusters. These clusters then become the condensation cores for silicate material at even larger distances from the star where the bulk of dust is formed. If we were to observe such an idealized envelope in spectral lines, we would see a hot (say $>$1400\,K) layer of gaseous TiO surrounding the stellar photosphere. It would be embedded in a more extended layer of TiO$_2$ gas characterized by lower temperatures, of $\sim$1000\,K. Spectroscopic signatures of both oxides should disappear at distances larger than 3\,$R_{\star}$, if both species are effectively depleted to form dust. Since the stellar atmospheres of our interest are very much influenced by (pulsation) shocks (e.g. \cite{Liljegren2016}), the real dust-formation zones are infinitely more complex, especially when it comes to chemical processes involved (\cite{Cherchneff2006,Cherchneff2012,Cherchneff2013,Gobrecht2016}). However, the nucleation scenario sketched here serves as an illustration of the idea how observations of gas phase molecules can help us understand the dust formation in stars.

About a decade ago, we did not have good direct access to tracers of any of the potential dust precursors in circumstellar gaseous media. A few stars had been known to exhibit circumstellar \textit{emission} in optical electronic bands of TiO and AlO but these bands alone were difficult to interpret and observations were scarce (\cite{KaminskiAlOVY}). Seeking the understanding of circumstellar nucleation gave a strong incentive to look for molecular dust precursors especially at submillimeter (submm) wavelengths because (I) pure rotational spectra of simple molecules lead to unambiguous identification of the carriers and (II) with the advent of interferometric arrays with unprecedented angular resolution \textit{and} sensitivity, especially ALMA, we are able to spatially resolve the regions where first solids form. Fortunately, such molecular submillimeter tracers were found in the last few years. The first such observations were acquired towards the iconic red supergiant VY\,CMa.

\section{Titanium oxides in the red supergiant VY\,CMa and AGB stars}
% firts detection
VY\,CMa is the prime source for first-detection experiments because with an extreme mass-loss rate of 10$^{-4}$\,M$_{\odot}$\,yr$^{-1}$ it accumulated a massive ($\sim$0.1--1.0\,M$_{\odot}$) envelope rich in molecules and dust. The circumstellar cloud of VY\,CMa has a characteristically complex substructure seen at optical to millimeter wavelengths (\cite{Humphreys2007}), which is most likely related to huge convective cells on the surface of this supergiant. The envelope was observed in an interferometric line survey obtained with the Submillimeter Array (SMA; \cite{KamiSurvey}). For each channel in the 279.1--355.1 GHz range, a high-sensitivity map at a resolution of about 1 arcsec, corresponding to 1200\,AU, was produced, sufficient to resolve emission from most molecules in the envelope. Among the nineteen identified molecules were TiO and TiO$_2$. This was the first detection of pure rotational emission of TiO in an astronomical object at submillimeter wavelengths and of TiO$_2$ at any wavelength (\cite{KamiVYTi}). The detection was later confirmed with the Plateau de Bure Interferometer (PdBI).

\begin{figure}[t]
  \begin{minipage}[c]{0.53\textwidth}
    \includegraphics[width=\textwidth]{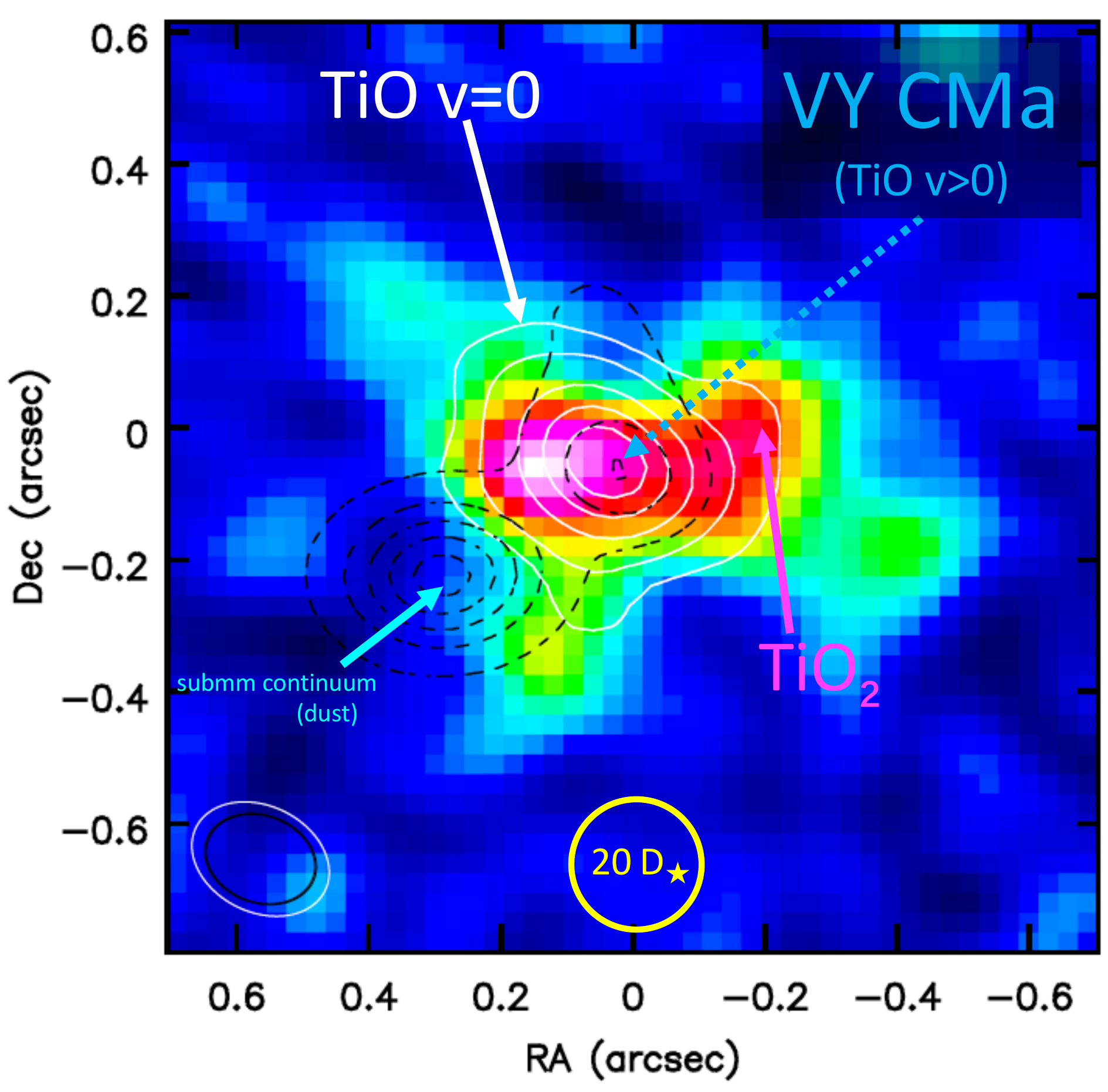}
  \end{minipage}\hfill
  \begin{minipage}[c]{0.45\textwidth}
    \caption{Emission of titanium oxides around the red supergiant VY\,CMa as seen by ALMA in 2015 (Kami\'nski et al., in prep.). The position of the star is marked with a dashed arrow and it is the same position where emission of TiO from excited vibrational states was found by Gottlieb et al. (in prep.). Solid contours show the emission of TiO and dashed contours show the submillimeter continuum emission of dust. The background image represents TiO$_2$ emission.} \label{figVY}
  \end{minipage}
\end{figure}

% SMA results for VY
Observing both of these oxides in rotational lines was very encouraging, giving us hope that we can study dust nucleation in stars using submillimeter interferometers. The first observations of VY\,CMa showed already some interesting features. The emission of TiO$_2$ was resolved and could be traced out to a radius of about 50\,$R_{\star}$. The excitation temperature of the TiO$_2$ gas was low, of 225\,K only. By combining spectra from SMA, PdBI, and UVES at the Very Large Telescope, Kami{\'n}ski et al. derived the extent of the TiO emission to be 30\,$R_{\star}$ and an excitation temperature of about 600\,K. These observational characteristics were in stark contrast to the simple prediction outlined here in Sect.\,\ref{S2}: both titanium oxides are abundant in the outer, cool parts of the envelope where the dust is already fully formed. This readily indicated that, at least in the violent environment of a supergiant such as VY\,CMa, titanium oxides do not play a major role in dust nucleation (alternatively, shock sputtering may bring back titanium oxides to the gas phase in the outer envelope; see \cite{KamiVYTi}).    

% update on ALMA obs.
Eventually, VY\,CMa was observed in lines of TiO and TiO$_2$ with ALMA (\cite{DeBeck}; Kami\'nski et al., in prep). As shown in Fig.\,\ref{figVY}, the ALMA maps expose both emission regions in great detail.  In particular, they show the transition from monoxide- to dioxide-dominated regions. Additionally, it was recently recognized that the SMA survey spectra contain emission features of TiO from vibrationally excites states ($v>0$; Gottlieb et al., in prep.) giving us an unprecedented view on the gas-phase titanium chemistry throughout the envelope. All these recent observations reinforce the conclusion that titanium oxides are unlikely dust precursors in VY\,CMa. 
%It is however unlikely that we will have a better view on both oxides in any other star any time soon and thus these observations are invaluable for astrochemistry studies in this important source.

\section{Titanium oxides in AGB stars}
\begin{figure}[t]
% \vspace*{-2.0 cm}
\begin{center}
 \includegraphics[width=\textwidth]{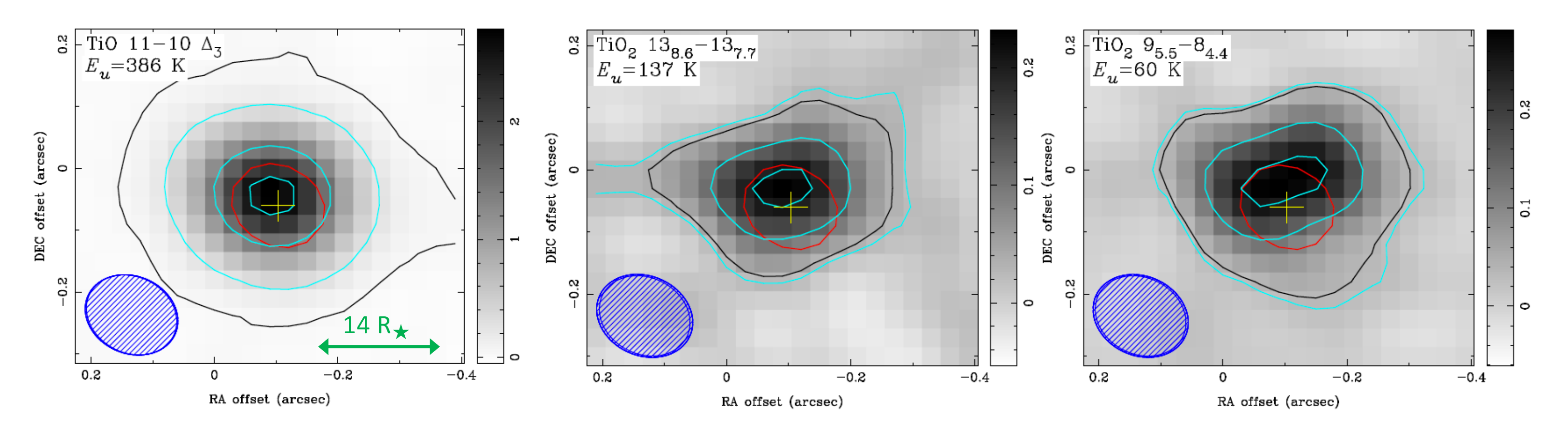}
% \vspace*{-1.0 cm}
 \caption{Emission of titanium oxides in Mira as seen by ALMA in 2015 (\cite{KamiMiraTi}). Left panel shows emission of TiO and the two other panels show emission of TiO$_2$ in the transitions indicated in each panel.The cross marks the position of the star and the associated ellipse represents the extent of the beam-smeared radio-photosphere. Light contours show molecular emission at 10, 50, and 90\% of the peak and the dark contours represent the 3$\times$rms noise levels of each map. ALMA beams are also shown (lower left corners).}
   \label{FigMiraTi}
\end{center}
\end{figure}

Soon after the first detection of TiO and TiO$_2$ in VY\,CMa, both oxides were detected in three M-type AGB stars: $o$\,Ceti (\cite{KamiMiraTi}), R\,Dor and IK\,Tau (\cite{DecinSurvey}). Mira was observed in the greatest detail with multiple instruments, including APEX, {\it Herschel}, and ALMA. Emission of both oxides was resolved by ALMA -- see Fig.\,\ref{FigMiraTi}. The emission of TiO is rather compact, homogeneous, and well centered on the position of the star. We can trace its emission in a region with an $e$-fording radius of about 4\,$R_{\star}$ and a typical excitation temperature of 470\,K. The emission of TiO$_2$ is more patchy (or clumpy), not centered on the stellar position, and slightly more extended than TiO with a Gaussian radius of 5.5\,$R_{\star}$; the excitation temperature of TiO$_2$ is lower, 174\,K. Optical spectra additionally reveal the presence of atomic Ti within a few $R_{\star}$. It actually appears that we trace nearly all of the elemental titanium available within these few $R_{\star}$ --- the derived molecular abundances are very close to the elemental abundance of Ti. With such high abundances of gaseous titanium oxides, it is very unlikely that they provide the first seeds for dust formation in Mira. The theoretical studies emphasize that nucleation starting from titanium oxides requires very efficient condensation of these Ti species, which is clearly not the case in Mira, even if we consider all uncertainties involved in deriving the molecular abundances.

That titanium oxides do not play the major role in dust nucleation was concluded before for VY\,CMa. \cite{DecinSurvey} arrived to a similar conclusion after analyzing both oxides in R\,Dor and IK\,Tau in ALMA maps. Generalizing, it is therefore very likely that in most evolved O-rich stars the formation of solids does not start with titanium oxides but with some other compounds. This result is surprising because it is widely believed that titanium is almost completely depleted from the gas phase in the interstellar medium and there are certainly presolar AGB grains that contain titanium oxides. Perhaps titanium and titanium oxides are incorporated into grains in the region where silicates form (\cite{KamiMiraTi}). 
%Even though they are not the major player in forming first hot solids, Ti-bearing species may be incorporated into grains at larger distances from the star (\cite{KamiMiraTi}). 

\section{Alumina dust precursors in VY\,CMa?}
Alumina seeds, if present in circumstellar envelopes, were proposed to be composed mainly of Al$_2$O$_3$ clusters (\cite{Gobrecht2016}). It is however unclear what the primitive gas precursors of these solids are (\cite{Alvarez2016}); most studies \textit{assume} it is the simplest oxide, i.e. AlO, and observational efforts focused mainly on this molecule, too. The first millimeter-wave observation of rotational transitions of AlO was obtained over a decade ago by \cite{ZiurysAlO} and the detection experiment was again conducted towards the red supergiant VY\,CMa. Other Al-bearing molecules were also observed in its complex envelope, including AlOH (\cite{ZiurysAlOH}) and AlCl (\cite{KamiSurvey}). Although AlO is easiest to observe owing to a combination of abundance and partition-function effects, it is actually AlOH that is the most abundant carrier of Al in VY\,CMa. These early studies found that the gas-phase molecular carriers of Al spread to large distances from the star, i.e. out to tens of $R_{\star}$. The bulk of the gas is therefore not directly related to the dust formation close to the star. 

\begin{figure}[t]
  \begin{minipage}[c]{0.53\textwidth}
    \includegraphics[width=\textwidth]{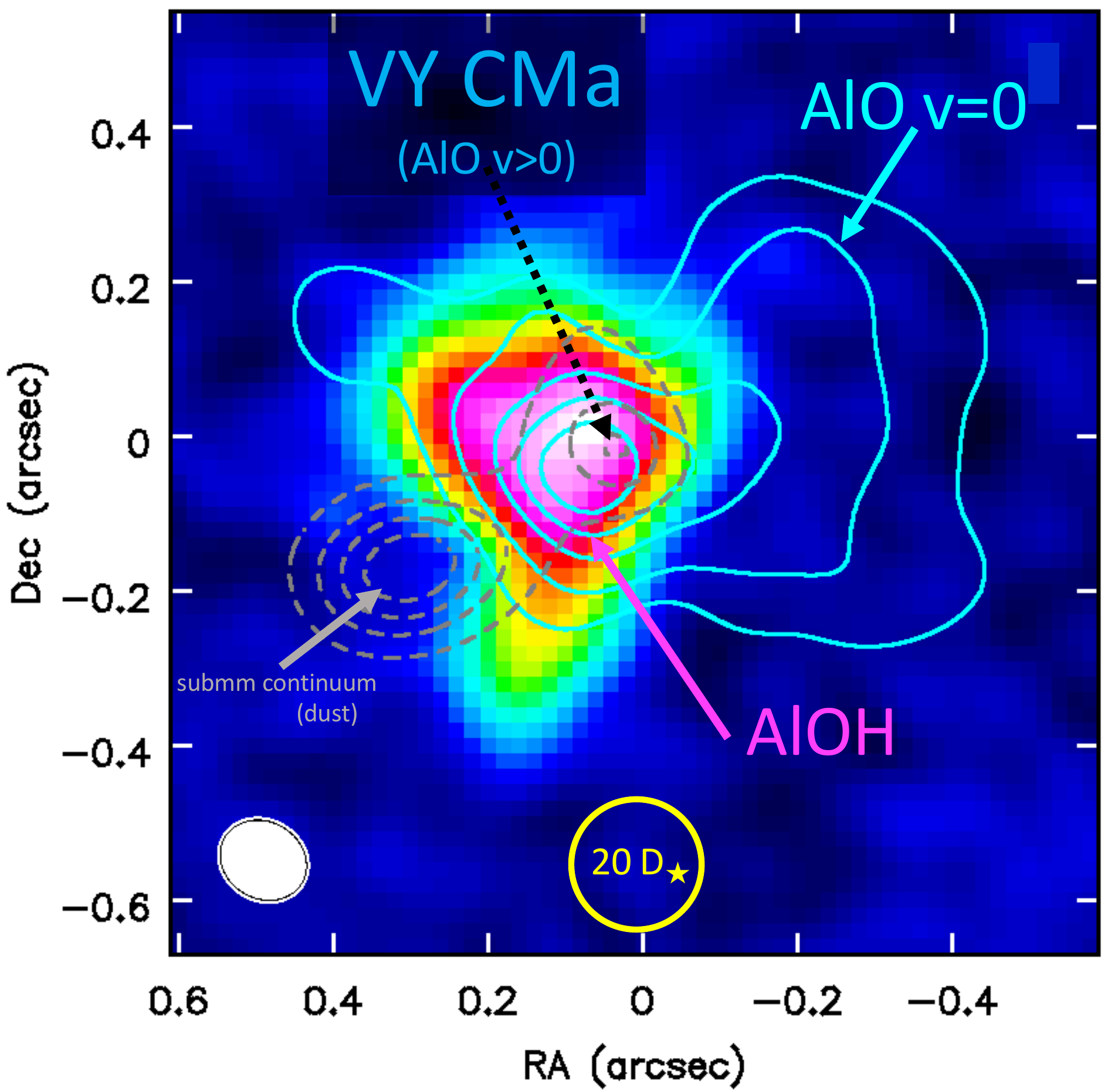}
  \end{minipage}\hfill
  \begin{minipage}[c]{0.45\textwidth}
    \caption{Emission of aluminium oxides around the red supergiant VY\,CMa as seen by ALMA in 2015 (Kami\'nski et al., in prep.). The position of the star is marked with a dashed arrow and it is the same position where emission of AlO from excited vibrational states was found by Gottlieb et al. (in prep.). Solid contours show the emission of AlO and dashed contours show the submillimeter continuum emission of dust. The background image represents AlOH emission.} \label{figVYAlO}
  \end{minipage}
\end{figure}

Since these pioneering observations, we improved the observational basis of Al-bearing species in VY\,CMa. The $B-X$ electronic bands of AlO in the optical was analyzed in detail in \cite{KaminskiAlOVY}. A range of pure rotational transitions of AlO were observed in wide band surveys with \textit{Herschel} and the SMA (\cite{SurveyHerschel}; \cite{KamiSurvey}); most recently, Gottlieb et al. (in prep.) were able to recognize pure rotational lines from excited vibrational states of AlO. Very detailed maps of the AlO and AlOH emission regions were obtained with ALMA at $\sim$0.1 arcsec resolution (Kami\'nski et al., in prep.) showing very intricate distributions of the molecular emission with respect to each other and with respect to the location of dust clumps and the stellar photosphere (see Fig.\,\ref{figVYAlO}). It appears that the gas-phase chemistry of Al-bearing species in this supergiant is particularly complex and the relation of the rich observational material to the state-of-the-art chemical models of AGB stars (\cite{Gobrecht2016}) is unclear and perplexing. There is hope that Al chemistry in ''better-behaving" AGB stars is going to be easier to interpret and indeed will allow us to shed more light on the alumina-dust formation.  

\section{AlO and AlOH as the alumina-dust precursors in AGB stars}
\begin{figure}[t]
% \vspace*{-2.0 cm}
\begin{center}
 \includegraphics[width=\textwidth]{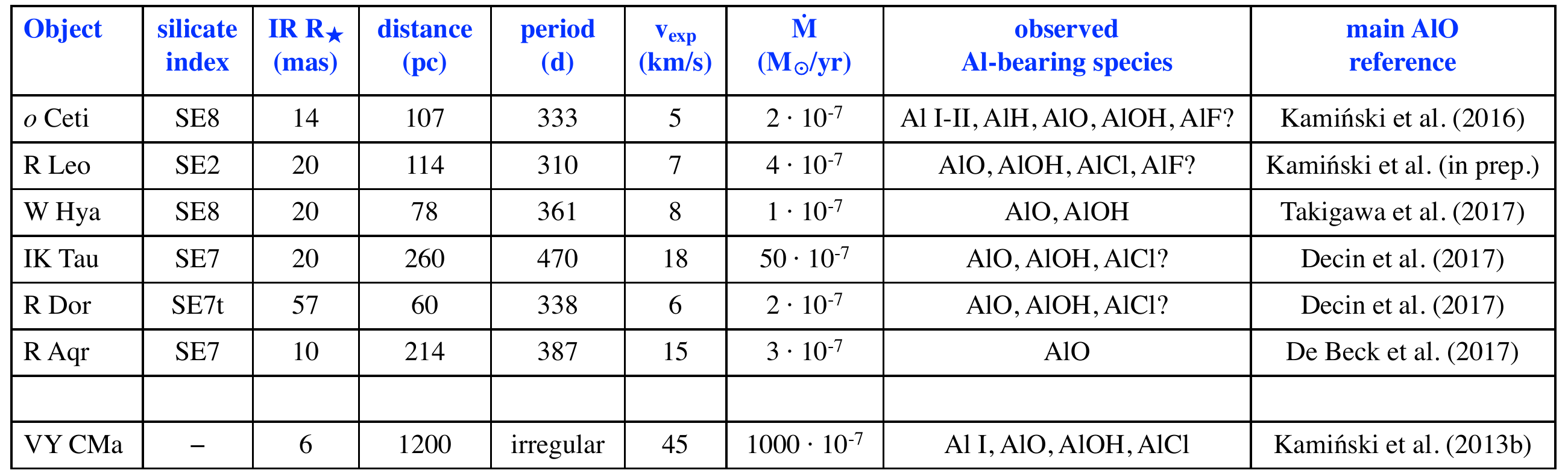}
% \vspace*{-1.0 cm}
 \caption{A list of evolved oxygen-rich stars observed in Al-bearing species at (sub-)millimeter wavelengths. The main reference reporting the most relevant data is listed in the last column. The stellar and envelope properties listed here are only approximate and were compiled from a variety of sources. The ''silicate indexes" are from \cite{SloanPrice1998}.}
   \label{FigTable}
\end{center}
\end{figure}

Encouraged by the successful detection experiments in VY\,CMa, there has been several observational studies of gas-phase Al-bearing species in M-type AGB stars (\cite{KamiMiraAl,DeBeckAlO, Decin2017, Takigawa2017}). The sources with successful detection are listed and briefly characterized in Fig.\,\ref{FigTable}. Most were observed only in AlO but the number and depth of alumina nucleation studies is increasing fast. The list of detected AGB sources covers already a relatively wide range of envelope or stellar properties, including different mass-loss rates, pulsation periods, wind terminal velocities, and types of infrared spectra (as indicated by the silicate-sequence index of \cite{SloanPrice1998}). 

\begin{figure}[t]
% \vspace*{-2.0 cm}
\begin{center}
 \includegraphics[width=\textwidth]{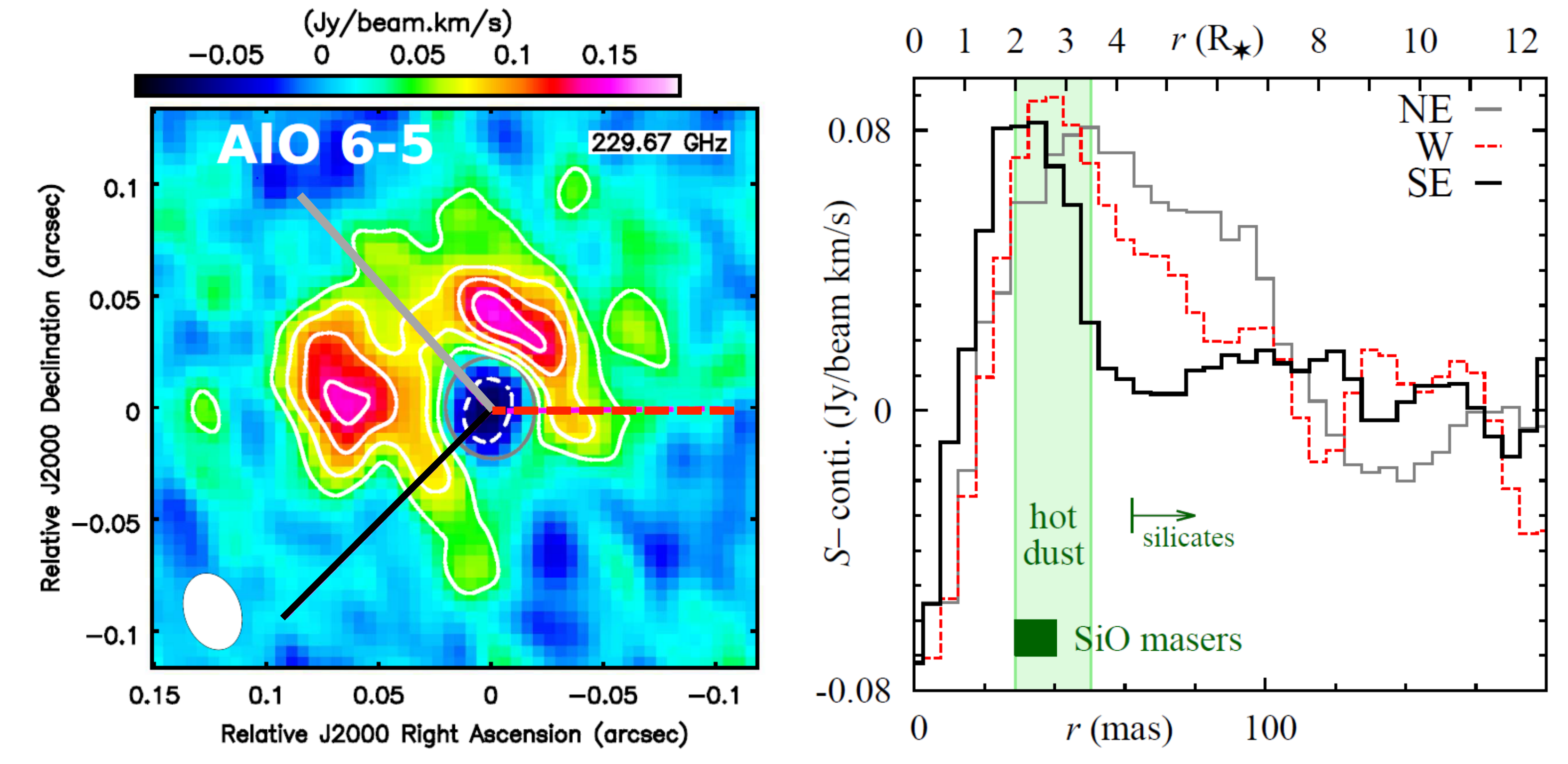}
% \vspace*{-1.0 cm}
 \caption{{\bf Left:} Emission of AlO in Mira as observed with ALMA (background image and contours). The stellar radio photosphere is resolved in these observations and its physical size is illustrated with an ellipse in the center. We observe AlO absorption (the dashed white contour) towards the stellar disk from cool gas located between us and the disk. The emission regions show that the distribution of warm AlO gas is very clumpy. {\bf Right:} Sample spatial profiles of AlO intensity along three cuts shown in the left panel. The drop in intensity at 2.0--3.5\,$R_{\star}$ coincides with the location of hot dust in Mira and may be a signature of AlO depletion into solids. Reproduced from \cite{KamiMiraAl}.}
   \label{FigMiraAlO}
\end{center}
\end{figure}

Among these sources, Mira was characterized in most detail and in the largest number of Al-bearing species. In particular, ALMA observations at a resolution of $\sim$30\,mas resolved the radio photosphere and the dust-forming region of Mira. The distribution of AlO emission mapped with ALMA is shown in Fig.\,\ref{FigMiraAlO} (\cite{KamiMiraAl}). The emission is particularly inhomogeneous or patchy. Spatial cuts, also shown in Fig.\,\ref{FigMiraAlO}, illustrate that the main emission region extends up to a radius of about 3.5\,$R_{\star}$ beyond which the emission drops significantly. Numerous lines of other species are observed in the same region and do not show such a drop. It was therefore suggested that the observed emission distribution reflects mainly changes in abundance rather than being an excitation effect. Since hot dust (of yet unknown mineralogy) has been directly observed in Mira at radii 2.0--3.5\,$R_{\star}$ in different pulsation phases, it is most tempting to interpret the drop in AlO abundance at $\sim$3.5\,$R_{\star}$ as a signature of depletion into first solids. However, there is currently no way of knowing whether AlO is converted (in a chain of reactions) to solid Al$_2$O$_3$ or to other Al-bearing gas-phase species. In the region where AlO and hot dust  are collocated, the abundance of AlO constitutes only 1--10\% of the total reservoir of elemental Al. This is fully consistent with effective condensation of alumina solids at 2.0--3.5\,$R_{\star}$ from AlO but, admittedly, other interpretations are possible. Based on observations at lower angular resolutions, \cite{Decin2017} derived similarly low abundances ($\sim$1\%) of AlO and AlOH in the envelopes of R\,Dor and IK\,Tau. Contrary to what was concluded above for seed formation from titanium oxides, it therefore appears that AlO or AlOH may well be the precursor species of the first solids in O-rich AGB stars.     

If indeed the clumpy AlO gas in Mira's envelope gives origin to the first solids, the distribution of hot dust should be clumpy too. This is currently investigated in Mira through infrared interferometic imaging with GRAVITY on the VLTI (K. T. Wong, priv. comm.). Moreover, the relative distributions of (polarized) continuum traced with VLT/SPHERE and rotational emission of AlO were also analyzed by Khouri et al. (see the contribution in this volume). Objects other than Mira were studied in this context as well. The AGB star W\,Hya was imaged at a high resolution with ALMA and its AlO emission shows similar features as these seen in Mira: it extends only out to $\sim$3\,$R_{\star}$ and shows strong inhomogenities with the azimuthal angle (\cite{Takigawa2017}). Fortunately, within 150 days of the ALMA observations, W\,Hya was observed with VLT/SPHERE in scattered light (\cite{Ohnaka2016}). These optical observations show remarkably similar spatial characteristics of dust and AlO emission. Takigawa et al. interpreted this striking resemblance as a signature of nucleation of AlO into solid alumina, but the similarity may simply reflect the overall density variations within the inner envelope. The relation between AlO (or AlOH) and the clumpy nature of the inner envelopes of AGB stars in indeed intriguing. 

% R LEO observations -skipped

\section{Alternative dust precursors}
Gas-phase molecules other than the Ti- and Al-bearing ones were investigated in circumstellar media in order to establish their role in inorganic dust formation, including NaCl (\cite{Milam}; \cite{DecinNaCl}), SiO (\cite{Wong}; \cite{Takigawa2017}), and FeO (a tentative detection of \cite{DecinFeO}). It appears they are, however, unlikely to be the major providers of seed particles in the majority of O-rich stars.

\section{Challenges in interpreting the data and in studying nucleation}
Despite the increasing number of observations of the potential gas-phase dust precursors, our understanding of chemistry and nucleation processes taking place close to the stellar photospheres is challenged by the complexity of the circumstellar environments and the physio-chemical processes involved. Taking Al-bearing species as an example, I offer below a brief overview of a few of the most complicating factors that often stop us from drawing firm conclusions from the observations.   

{\underline{\it Molecular tracers.}}
As noted above, most stars so far have been observed mainly in rotational lines of AlO from within the ground vibrational level. The number of stars with detected lines of AlOH is increasing (see Fig.\,\ref{FigTable}). However, in order to constrain the chemistry leading to nucleation, we should try to trace as many species as possible. For instance, the observed drop in abundance of AlO near 3\,$R_{\star}$ may be interpreted as  conversion of AlO into other gas-phase oxides, say to AlO$_2$, with no associated condensation taking place. Mira remains the source with largest number of species observed -- with features of Al\,I, Al\,II, AlH, AlO, AlOH, and AlF observed in different wavelength regimes (\cite{KamiMiraAl}, K. T. Wong, priv. comm.). Other observable species include AlS and AlCl but there are gas-phase molecules which we currently are not able to observe, including for instance Al$_2$, Al$_2$O, AlO$_2$, and Al$_2$O$_2$. Some of them cannot be observed owing to very weak or lack of rotational transitions; for others, spectroscopic data are incomplete or missing. Because we are interested in envelope regions close to the star, where the prevailing conditions lead to high excitation, the observations should include also electronic, ro-vibrational, and high-$J$ and high-$v$ rotational spectra. The already mentioned identification of rotational lines of AlO and TiO at $v>0$ levels (Gottlieb et al., in prep.) will certainly add to our understanding of the inner, dust-forming regions of the envelopes.    

{\underline{\it Variability.}}
Observations of Mira in Al- and Ti-bearing molecules show a very high level of variability. Optical spectra of M-type Mira stars have long been  known for their erratic changes in molecular bands of AlO and TiO (\cite{Garrison}). Decades of optical observations of circumstellar features of AlO and Al\,I in $o$\,Ceti (\cite{KamiMiraAl}) do not show any regularity in the intensity changes over the pulsation cycle or on any longer time scales. Multiepoch ALMA observation of the same source in pure rotational transitions also show such an uncorrelated variability. This behavior is most likely caused by the stochastic nature of the pulsation shocks that strongly influence this part of the circumstellar envelope   (\cite{Cherchneff2006,Gobrecht2016,Liljegren2016,Liljegren2018}). The variability is a major obstacle in determining the physical conditions in the gas since it requires nearly-simultaneous observations in multiple transitions. Such coordinated observations are difficult to arrange, especially when  different observing facilities need to be involved. Of course, studying variability in different lines of many different species is also very expensive in terms of telescope time. Similarly, we require coordinated observations of dust tracers, such as optical observation of the (polarized) scattered light and mid-infrared observations of the spectral signatures of solids.

{\underline{\it Non-LTE gas excitation.}}
Although optical electronic bands of relevant circumstellar molecules could be observed for many M-type stars (\cite{Keenan}), they are currently of limited use because there are no straightforward ways to derive reliable gas characteristics from the resonantly scattered emission bands (but see \cite{KaminskiAlOVY}). Most of the studies thus far have therefore used pure rotational transitions observed at millimeter wavelengths as the diagnostics of gas physical conditions and molecular abundances. The radiative transfer problem for these lines has been solved at different levels of sophistication but all studies ignored the 3-dimensional complexity of the emitting regions (clumping), the influence of variable stellar radiation (radiative pumping), and the influence of shocks which most likely introduce non-LTE effects to molecular excitation. This introduces many uncertainties in the derived column densities. Additionally, without a reliable tracer of the local hydrogen densities, the derived absolute abundances or abundance profiles are highly unreliable. Dust or CO observations are used as a proxy of hydrogen densities but the required conversion factors are also poorly known for individual objects and especially in this very dynamic part of the envelope where dust has not fully formed and active chemistry is taking place. A lot of effort is necessary to build more reliable tools that would translate the observed line intensities into molecular abundances. After forcefully addressing these challenges we will truly be able to trace the depletion of molecules into solids.   

\medskip

{\small \noindent{\bf Acknowledgments.} I would like to thank my collaborators K. M. Menten, I. Cherchneff, N. Patel, and J. M. Winters for  providing feedback on an early version of this manuscript.}

%\begin{discussion}

%\discuss{Massey}{I'm wondering if you have considered the expected intrinsic dispersion in absolute magnitude of WRs -- if you consider the (large) mass range that becomes an early WN or late WC according to the evolutionary models, wouldn't you expect a large dispersion in M$_v$?}

%\end{discussion}

\end{document}